\documentclass[preprint,showpacs,preprintnumbers,chaos]{revtex4}
\usepackage{graphicx}

\begin{document}

\title{Long-term memory contribution as applied to the motion
of discrete dynamical systems}
\author{\ A.A. Stanislavsky\\
\textit{{\small Institute~of~Radio~Astronomy,
Ukrainian~National~Academy~
of~Sciences}}\\\textit{{\small4~Chervonopraporna~St.,
Kharkov~61002, Ukraine}} }

\begin{abstract}
We consider the evolution of logistic maps under long-term memory.
The memory effects are characterized by one parameter $\alpha$. If
it equals to zero, any memory is absent. This leads to the
ordinary discrete dynamical systems. For $\alpha$\ =\ 1 the memory
becomes full, and each subsequent state of the corresponding
discrete system accumulates all past states with the same weight
just as the ordinary integral of first order does in the
continuous space. The case with $0<\alpha<1$ has the long-term
memory effects. The characteristic features are also observed for
the fractional integral depending on time, and the parameter
$\alpha$ is equivalent to the order index of fractional integral.
We study the evolution of the bifurcation diagram among $\alpha$\
=\ 0 and $\alpha$\ =\ 0.15\,. The main result of this work is that
the long-term memory effects make difficulties for developing the
chaos motion in such logistic maps. The parameter $\alpha$
resembles a governing parameter for the bifurcation diagram. For
$\alpha>$\ 0.15 the memory effects win over chaos.
\end{abstract}


\maketitle

\section*{}\thispagestyle{empty}
{\bf The treatment of nonlinear dynamics in terms of discrete maps
(difference equations produced by numerical methods) is a very
important step in studying the qualitative behaviour of continuous
systems described by differential equations. The logistic map
represents one of the most important examples of an
one-dimensional discrete nonlinear map with the bifurcation
scenario well known. The complicated behaviour exhibited by the
logistic map is typical for a whole class of dynamical systems.
Fractional calculus occupies an appreciable place in order to
describe various kinds of wave propagation in complex media,
fractional kinetics of Hamiltonian systems, anomalous diffusion
and relaxation, etc. The fractional operator is a natural
generalization of the ordinary differentiation and integration.
When the operator depends on time, it is characterized by
long-term memory effects. The effects correspond to intrinsic
dissipative processes in the physical systems. In the application
to discrete maps this means that their present state evolution
depends on all past states with a power weight. Appearance of
long-term memory effects in the logistic map makes the coupling
among states stronger. This feature is plainly directed against
the development of the chaotic dynamics.}

\section{Introduction}
\thispagestyle{empty} The evolution of discrete dynamical systems
is described by the variables measured in discrete time steps. The
behavior of such systems is governed by return maps. They connect
the $(n+1)$th value of variables with the preceding $n$th value of
the variables through a functional dependence. A discrete
dynamical system can demonstrate a complex development even in
simple systems with one variable \cite{schuster}. If the system
falls into a series of bifurcations, it exhibits a transition from
a variety of periodic cycles to chaos.

A statistical treatment of the macroscopic equation of motion
leads to memory effects in dynamical Onsager coefficients, if
Markovian-like approximations are not admissible \cite{risken}.
The application of memory effects to the discrete systems
signifies that their dynamics at time $t+1$ will depend only not
on time $t$, but on former times $t-1,t-2,t-3$ and so on. Here
time is equivalent to the number of step. Therefore, $n+1$ can be
substituted for $t+1$ as well as $n$ can be $t$, and $t-1,
t-2,\dots$ can be replaced by $n-1, n-2,\dots$ respectively. The
introduction of a retardation of the linear term in the logistic
equation modifies the Feigenbaum scenario \cite{fick}. In this
case the periodic orbits are shifted. Several forms of nonlinear
maps with memory were also considered by a formal way in
\cite{fulinski}.

Last time the interest to the study of long-term memory effects
and fractional kinetics in physical systems increased too much
\cite{mk,psw,zaslavsky,zaslavsky1}. The fractional operator with
respect to time is characterized by long-term memory effects,
whereas one with respect to coordinates possesses non-local
(long-range) effects \cite{tz,lz}. The direct relationship among
the long-term memory effects, fractional calculus and the stable
distributions from the theory of probability has been established
in \cite{stanislavsky1,meerschaert}. Then it was shown \cite{zse}
that perturbed by a periodic force, the nonlinear oscillator with
fractional derivative exhibits a chaotic motion called the
fractional chaotic attractor. Although the fractional nonlinear
oscillator behaves like the stochastic attractor in phase space,
being periodically perturbed, the role of the polynomial
dissipation leads to a degradation of the attractor structure. The
aim of this work is to present the analysis of the influence of
long-term memory effects on the behavior of discrete systems.

The paper starts with a discrete conception of long-term memory
effects (Section~\ref{par2}). They are characterized by the
parameter similar to the order index of fractional integral. By
means of the computer simulation of bifurcation diagrams for the
quadratic and triangular maps (Section~\ref{par3}) with the memory
contribution, we demonstrates how the memory effects strangle the
chaotic motion of discrete dynamical systems.

\section{Mathematical description of long-term memory}\label{par2}

The mapping $x_{n+1}=f(x_n)$ does not have any memory, as the
value $x_{n+1}$ only depends on $x_n$. The introduction of memory
means that the discrete value $x_{n+1}$ is connected with the
previous values $x_n$, $x_{n-1}$, $x_{n-2}$, $\dots$, $x_1$.
Particularly, any discrete system will have a full memory, if each
state of the system is a simple sum of all previous states:
\begin{displaymath}
x_{n+1}=\sum^n_{i=1}f(x_i),
\end{displaymath}
where $f(x)$ is an arbitrary function suitable for the discrete
map. However, the above expression may tend to infinity because of
the sum accumulating all the values. In order to have a fixed
point in this mapping, the expression should be normalized as
\begin{equation}
x_{n+1}=\frac{1}{n}\sum^n_{i=1}f(x_i)\,.\label{eq1}
\end{equation}
When $p$ is a fixed point, the map (\ref{eq1}) gives
\begin{displaymath}
x_{n+1}=\frac{1}{n}\sum^n_{i=1}p=p.
\end{displaymath}
The starting conjection is that the map under long-term memory is
expressed in terms of
\begin{displaymath}
x_{n+1}=\frac{1}{n^\alpha}\sum^n_{i=1}b_if(x_i),
\end{displaymath}
where the weights $b_i$ and the parameter $\alpha$ characterize
the non-ideal memory effects.

Let the weights take the form
\begin{displaymath}
c_{i}^{(n)}=\cases{(1+\alpha)n^\alpha-
n^{\alpha+1}+(n-1)^{\alpha+1}\,,&if $i=0$;\cr 1\,,&if $i=n$;\cr
(n-i+1)^{\alpha+1}-2(n-i)^{\alpha+1}+ & \cr
+(n-i-1)^{\alpha+1}\,,& if $0<i<n-1$.\cr}
\end{displaymath}
In this connection it should be pointed out that the similar
representation is used for the robast algorithm of numerical
fractional integration \cite{dffl}. It has found an interesting
application for the analysis of the nonlinear reaction with
fractional dynamics \cite{stanislavsky2} as well as for the study
of fractional oscillations \cite{stanislavsky3}. Then the
normalized mapping is written as
\begin{equation}
x_{n+1}=\frac{\sum_{i=0}^nc_{i}^{(n)}f(x_i)}{\sum_{i=0}^nc_{i}^{(n)}}
\,.\label{eq2}
\end{equation}
Here the sum starts with $i=0$ (though it could be realized from
$i=1$) by reason of the simplicity in writing the weights
$c_{i}^{(n)}$. To calculate the normalization factor presents no
difficulty, namely
\begin{displaymath}
\sum_{i=0}^nc_{i}^{(n)}=(1+\alpha)n^\alpha\,.
\end{displaymath}
It turns out that the factor is a power function just as one in
our foregoing conjection.

Now consider the following limit cases. If $\alpha=0$, the memory
is absent. Really, in this case all the weights are equal to zero
with the exception of $c_n^{(n)}=1$. Thus,  we come to an ordinary
map $x_{n+1}=f(x_n)$. For $\alpha=1$ the memory effects will be
full, namely
\begin{displaymath}
c_{i}^{(n)}=\cases{1\,,&if\quad $i=0$;\cr 1\,,&if\quad $i=n$;\cr
2\,,& if\quad $0<i<n-1$.\cr}
\end{displaymath}
Then this map becomes
$x_{n+1}=\frac{1}{n}[(f(x_0)+f(x_n))/2+\sum_{i=1}^{n-1}f(x_i)]$.
The full memory is ideal because each new state of the discrete
system sticks to the system's memory and has the same action upon
next states as all the others in memory. When $0<\alpha<1$, the
memory effects have a long-term dependence so that
\begin{equation}
x_{n+1}=\frac{1}{(1+\alpha)n^\alpha}\sum_{i=0}^nc_{i}^{(n)}f(x_i)
\,.\label{eq3}
\end{equation}
Certainly the memory is not ideal, as the contribution of more
early states is noticeably less than the contribution of following
ones on the present state of such systems. This is something like
a part of system states being lost. Recall that if a system does
not remember any previous state except for the present, it has no
memory effects.

In the next section we will show the results of numerical
simulations for the triangular and quadratic maps under the
long-term memory effects. This allows us to estimate their
influence on the evolution of the discrete systems. The memory
effects are governed by means of the parameter $\alpha$, whereas
the parameter of type $r$ serves as an order parameter for the
onset of chaos.

\section{Numerical results}\label{par3}
The triangular map is expressed in terms of the function
\begin{displaymath}
\Delta(x)=r\Bigl(1-2\Bigl|\frac{1}{2}-x\Bigr|\Bigr),
\end{displaymath}
and the quadratic map is written as
\begin{displaymath}
x_{n+1}=q x_n(1-x_n),
\end{displaymath}
where $r,\ q\ $ are constants. These logistic maps are very
popular in the theory of deterministic chaos. They help easily to
understand main features of discrete dynamical systems. In
particular, the quadratic map is the simplest nonlinear difference
equation, appears in many contexts, for example, a strongly damped
kicked rotator or the growth model of a population in a closed
area \cite{schuster}. For the triangular map with $r<1/2$ the
origin $x=0$ is the only stable fixed point to which all points in
the interval $[0\ 1]$ are attracted. The value $r$ plays a role of
``order parameter'', which indicates the onset of chaos. For
$r>1/2$ two unstable fixed points emerge. If the magnitude of the
parameter $r$ increases still more, then the information about the
position of a point in $[0\ 1]$ is lost. This may result in chaos.
The function $\Delta(x)$ is a simple model which for $r>1/2$
generates chaotic sequences $x_0, \Delta(x_0),
\Delta[\Delta(x_0)], \dots$, and due to its simple form, all
quantities in this chaotic state can be calculated explicitly. We
go the same way for the maps taking into account the long-term
memory effects.

\begin{figure}
\centering
\includegraphics[width=12 cm]{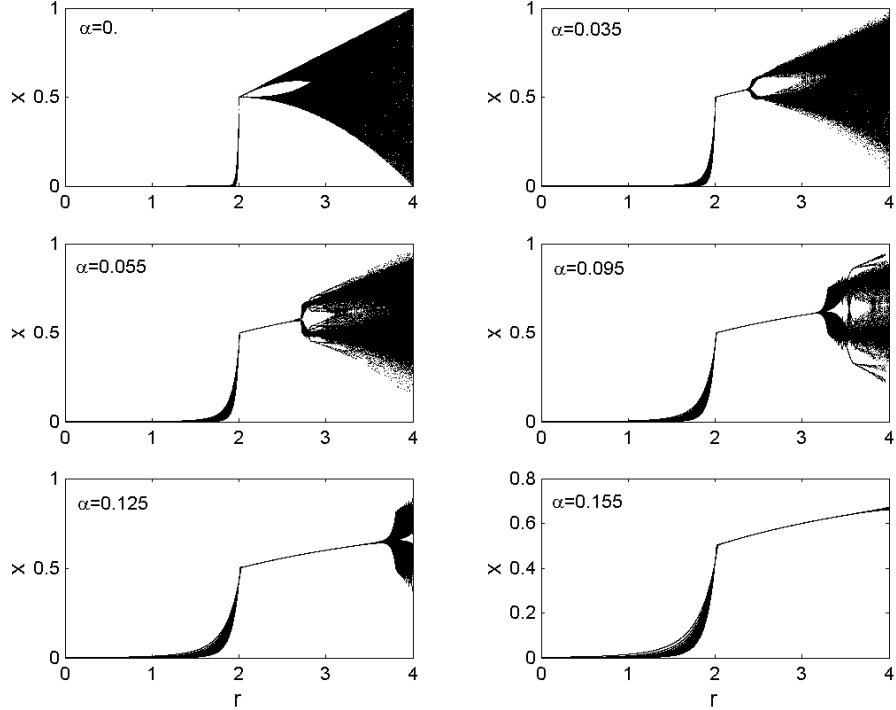}
\caption{\label{fig1}The bifurcation diagrams for the triangular
logistic map with long-term memory effects. The parameter $\alpha$
accounts for the memory contribution, the constant $r$ relates to
the onset of chaos for this map (all details in the text).}
\end{figure}

\begin{figure}
\centering
\includegraphics[width=12 cm]{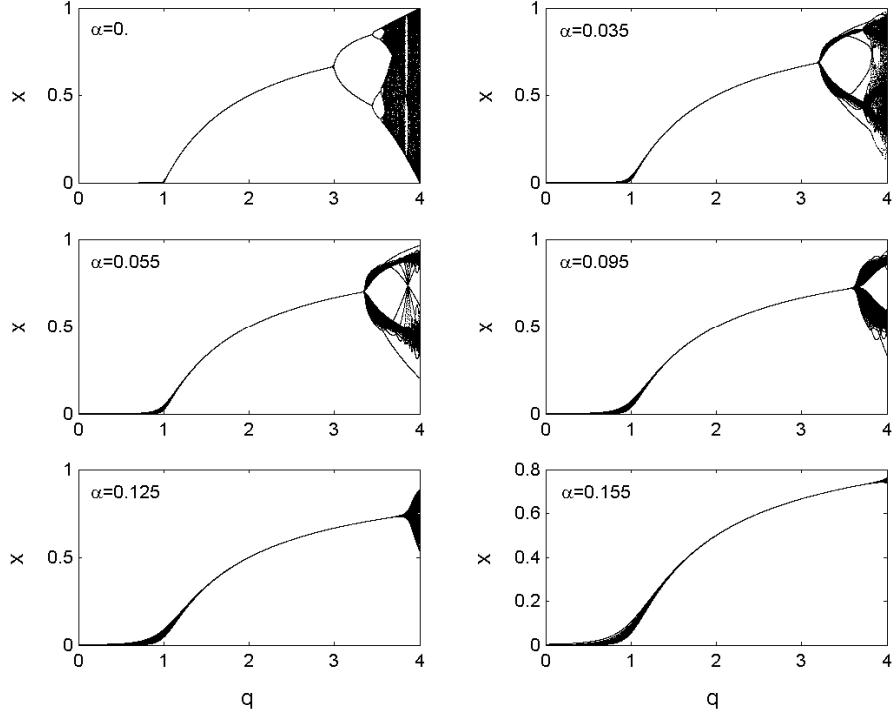}
\caption{\label{fig2}The bifurcation diagrams for the quadratic
logistic map with long-term memory effects. The parameter $\alpha$
characterizes the memory influence, the value $q$ is the constant
of the map (see details in the text).}
\end{figure}

The long-term memory distinguishes itself by a slow decay
contribution of former values for discrete maps. The procedure of
calculating the string of system states requires to evaluate in
the coefficients $c_{i}^{(n)}$. Therefore, the duration of
calculations notably increases with the growth of $n$. To remember
``the simple dynamical systems do not necessarily lead to simple
dynamical behavior'' \cite{may}, we guess that a weak memory
effects with $\alpha\to 0$ almost will not influence on the
complicated behavior of such discrete systems. By means of
numerical simulation we establish the evolution of the simple
logistic maps under long-term memory, when the parameter $\alpha$
tends to one. For this purpose the bifurcation diagram is
constructed. Figure \ref{fig1} shows a set of bifurcation diagrams
to the triangular map with various values $\alpha$. A number of
diagrams, represented in Fig.~\ref{fig2}, corresponds to the
quadratic case. The number of iterations $n$ equals to 3500. In
these figures it is seen that the memory effects step-by-step in
the growth of $\alpha$ block up the development of chaotic
behavior in the logistic maps. After $\alpha>0.155$ the evolution
of the given maps tends to a regular stable fixed point.

What can we say about a sequence of iterates if there are unstable
fixed points? The $n$th iterate $\Delta^n(x)$ is piecewise linear
and has the slope $\frac{d}{dx}\Delta^n(x)=2^n$, except for the
countable set of points $j\cdot 2^{-n}$ with $j=0, 1, \dots, 2^n$.
The Liapunov exponent measures the average loss of information
about the position of a point in $[0\ 1]$ after one iteration. For
the general triangular map under long-term memory effects the
Liapunov exponent becomes
\begin{eqnarray}
\lambda(x)&=&\lim_{n\to\infty}\frac{1}{n}\ln\Bigl|
\sum_{i=0}^nc_{i}^{(n)}\frac{d}{dx}\,\Delta^i(x)\Bigr|\
\leq\lim_{n\to\infty}\frac{1}{n}\ln\left\{\sum_{i=0}^nc_{i}^{(n)}
\Bigl|\frac{d}{dx}\,\Delta^i(x)\Bigr|\right\}\ \nonumber\\
&=&\lim_{n\to\infty}\frac{1}{n}\ln\Bigl\{\sum_{i=0}^nc_{i}^{(n)}
(2r)^i\Bigr\}=\lim_{n\to\infty}\frac{1}{n}\ln\Bigl\{(2r)^n
\sum_{i=0}^nc_{i}^{(n)}(2r)^{i-n}\Bigr\}\ \nonumber\\ &=&\ln(2r).
\nonumber
\end{eqnarray}
The equality $\lambda(x)=\ln(2r)$ holds true for $\alpha=0$, when
any memory effects are missing. If $0<\alpha<1$, then  the
Liapunov exponent decreases. Therefore, the chaotic states typical
for the ordinary triangular map die out with the growth of memory
effects, i.\ e.\ when the parameter $\alpha$ tends to one.

\section{Conclusions}\label{par4}
We have shown that if a discrete dynamical system is exposed to
the long-term memory effects, then the chaotic sequences which are
presented in the system without any memory are squeezed out. As
the memory mounts, chaos and its traces disappear. The magnitude
$\alpha_{\rm crit}\approx 0.155$ looks like a ``critical point''
that indicates a transition from chaos to order. It should be
noticed that the fine structure in the iterates for the maps is
just washed out because of the memory effects. Nevertheless, for
sufficiently small values of $\alpha$ there is still a remarkable
transition to chaos.

\section*{Acknowledgements}
The author acknowledges G.M. Zaslavsky for fruitful discussions,
M. Edelman for his help as well as the referees for their useful
remarks.

\end{document}